\documentclass[aps,prb,preprint,floats,showpacs]{revtex4}
\usepackage{epsfig,amssymb,amsfonts,amsmath}
\usepackage{graphicx}
\input{epsf}
\begin{document}

\title{Topological bands in two-dimensional networks of metamaterial elements}

\author{Vassilios Yannopapas}
\email{vyannop@upatras.gr} \affiliation{Department of Materials
Science, School of Natural Sciences, University of Patras,
GR-26504 Patras, Greece}

\date{\today}

\begin{abstract}
We show that topological frequency band structures emerge in
two-dimensional electromagnetic lattices of metamaterial
components without the application of an external magnetic field.
The topological nature of the band structure manifests itself by
the occurrence of exceptional points in the band structure or by
the emergence of one-way guided modes. Based on an EM network with
nearly flat frequency bands of nontrivial topology, we propose a
coupled-cavity lattice made of superconducting transmission lines
and cavity QED components which is described by the
Janes-Cummings-Hubbard model and can serve as simulator of the
fractional quantum Hall effect.
\end{abstract}

\pacs{42.70.Qs, 73.20.-r, 73.43.-f} \maketitle
\bibliographystyle{apsrev}

\section{Introduction}
The topological description of the quantum states of matter sets
in a new paradigm in the description and classification of atomic
solids. Namely, atomic solids whose energy band structure possess
nontrivial topological properties constitute a new class of
materials whose salient properties are robust to phase transitions
which modify the symmetry order of the atomic solid. Prominent
examples of such topological atomic solids are the integer/
fractional quantum Hall (I/FQHE) systems and the topological
insulators (TIs). Well-known examples of topological properties
are the existence of chiral edge states in QHE systems and the
presence of gapless surface states in TIs which are both immune to
order-disorder phase transitions.

The advent of artificial electromagnetic (EM) structures such as
photonic crystals and metamaterials has established over the years
a continuous conveyance of ideas and methods from atomic solids to
their EM counterparts. Quite naturally, the concept of topological
order has been adapted to photonic crystals starting with the QHE:
a two-dimensional (2D) lattice of gyromagnetic/ gyroelectric
cylinders is a system with broken time-reversal symmetry
\cite{haldane} with frequency bands characterized by a nonzero
Chern number, allowing for the emergence of unidirectional
(one-way) edge states \cite{one_way} in analogy with the chiral
edges states in QHE systems such as a 2D electron gas or graphene
nanoribbons under magnetic field. Anomalous QHE can also be
simulated with artificial chiral metamaterials of gyromagnetic
components. \cite{yannop_rashba} In TIs \cite{ti_papers} and
quantum spin Hall systems \cite{mele_2005} in 2D, the presence of
magnetic field is not prerequisite for the appearance of
topological electron states. In analogy with atomic TIs, in
certain 3D photonic crystals and metamaterials with proper design,
topological frequency bands appear without comprising
gyromagnetic/ gyroelectric materials which require the application
of external magnetic field in order to break time-reversal
symmetry. \cite{chen_pti,zhang_pti,yannop_pti}

In this Letter, we propose a class of 2D EM networks possessing
topological frequency bands without the application of an external
magnetic field. Namely, we show that topological bands emerge in
2D lattices of EM resonators connected with left- and right-handed
metamaterial elements such as transmission lines or waveguides
loaded with a negative refractive-index medium. The topological
nature of the corresponding frequency bands is manifested by the
emergence of an exceptional point for transverse electric (TE) and
by the generation of one-way modes for transverse magnetic (TM)
waves. In the latter case, the system can be viewed as a simulator
of the FQHE for polaritons.

\section{Lattice of coupled dipoles.}

The EM crystals under study here are amenable to a photonic
tight-binding description within the framework of the
coupled-dipole method. \cite{cde} The latter is an exact means of
solving Maxwell's equations in the presence of nonmagnetic
scatterers. We consider a lattice of cavities within a lossless
metallic host. The $i$-th cavity is represented by a dipole of
moment ${\bf P}_{i}=(P_{i;x},P_{i;y},P_{i;z})$ which stems from an
incident electric field ${\bf E}^{inc}$ and the field which is
scattered by all the other cavities of the lattice. This way the
dipole moments of all the cavities are coupled to each other and
to the external field leading to the coupled-dipole equation
\begin{equation}
{\bf P}_{i}=\alpha_{i}(\omega) [{\bf E}^{inc} + \sum_{i' \neq i}
{\bf G}_{i i'}(\omega) {\bf P}_{i'}]. \label{eq:cde}
\end{equation}
${\bf G}_{i i'}(\omega)$ is the electric part of the free-space
Green's tensor and $\alpha_{i}(\omega)$ is the polarizability the
$i$-th cavity. Eq.~(\ref{eq:cde}) is a $3N \times 3N$ linear
system of equations where $N$ is the number of cavities of the
system.

For a particle/cavity of electric permittivity $\epsilon$ embedded
within a material host of permittivity $\epsilon_{h}$, the
polarizability $\alpha$ is provided by the Clausius-Mossotti
formula $\alpha=(3V/4 \pi) (\epsilon-\epsilon_{h})/(\epsilon +
2\epsilon_{h})$, where $V$ is the volume of the particle/ cavity.
For a lossless Drude-type (metallic) host i.e.,
$\epsilon_{h}(\omega)=1-\omega_{p}^{2} / \omega^{2}$ (where
$\omega_{p}$ is the bulk plasma frequency), the polarizability
$\alpha$ exhibits a pole at $\omega_{0}=\omega_{p} \sqrt{2/
(\epsilon +2)}$ (surface plasmon resonance). By making a Laurent
expansion of $\alpha$ around $\omega_{0}$ and keeping the leading
term, \cite{yannop_pti} we may write $\alpha= F/(\omega -
\omega_0) \equiv F/\Omega$ where $F=(27V/8 \pi)\omega_0 \epsilon/
(2 \epsilon+4)$. For sufficiently high value of the permittivity
of the dielectric cavity the electric field of the surface plasmon
is much localized at the surface of the cavity. As a result, in a
periodic lattice of cavities, the interaction of neighboring
surface plasmons is very weak leading to much narrow frequency
bands. By treating such a lattice in a tight binding (TB) manner,
\cite{yannop_pti} we may assume that the Green's tensor ${\bf
G}_{i i'}(\omega)$ does not vary much with frequency and
therefore, ${\bf G}_{i i'}(\omega) \simeq {\bf G}_{i
i'}(\omega_0)$. In this case, Eq.~(\ref{eq:cde}) becomes an
eigenvalue problem
\begin{equation}
\sum_{i' \neq i} {\bf G}_{i i'}(\omega_0) {\bf P}_{i'}= \Omega
{\bf P}_{i}. \label{eq:cde_eigen}
\end{equation}
where $F$ has been absorbed within the definition of $ {\bf G}_{i
i'}(\omega_0)$ and we have set ${\bf E}^{inc}={\bf 0}$ in
Eq.~(\ref{eq:cde}) as we are seeking the eigenmodes of the system
of cavities.  In the following, we will be dealing with 2D
lattices of cavities. We can, therefore, treat separately the case
where the electric field lies within the plane of cavities (TE
modes) from the case where the electric field is perpendicular to
the plane (TM modes).

\subsection{TE modes}

In this case, ${\bf P}_{i}=(P_{i;x},P_{i;y})$ and the Green's
tensor ${\bf G}_{i i'}(\omega_0)$ is given by
\begin{eqnarray}
{\bf G}_{i i'}(\omega_0)= F q_{0}^{3} \Bigl[ C(q_{0} | r_{ii'}|)
{\bf
I}_{2} + J(q_{0} | r_{ii'}|) \left(%
\begin{array}{cc}
  \frac{x_{ii'}^2}{r_{ii'}^{2}} & \frac{x_{ii'}y_{ii'}}{r_{ii'}^{2}} \\
  \frac{x_{ii'}y_{ii'}}{r_{ii'}^{2}} & \frac{y_{ii'}^2}{r_{ii'}^{2}} \\
\end{array}%
\right) \Bigr]. \nonumber \\ \label{eq:g_tensor}
\end{eqnarray}
with ${\bf r}_{ii'}={\bf r}_{i}-{\bf r}_{i'}$, $q_{0}=\omega_0
\sqrt{\epsilon_{h}(\omega_0)}/c$ and ${\bf I}_{2}$ is the $2
\times 2$ unit matrix. Since we focus our attention around the
surface plasmon frequency $\omega_0$, we operate in the
subwavelength regime where $q_{0} | r_{ii'}| \ll 1$. In this
regime, the functions $C(q_{0} | r_{ii'}|)$, $J(q_{0} | r_{ii'}|)$
are written as
\begin{eqnarray}
 q_{0}^{2} F C(q_{\parallel} | r_{ii'}|)&\simeq& - q_{0}^{3} F
J(q_{\parallel} | r_{ii'}|) \simeq q_{0}^{3} F \exp(i q_0 |
r_{ii'}|)/ (q_0 |r_{ii'}|) \nonumber \\ &=& t_{ii'} \exp(i
\phi_{ii'}) \label{eq:tb_par}
\end{eqnarray}
where $t_{ii'}$ and $\phi_{ii'}$ are real numbers. In what
follows, the cavities are connected via coupling elements, i.e.,
waveguides or transmission lines, in which case the phase factors
$\phi_{ij}$ are not necessarily related with the wavevector of the
host medium $\epsilon_{h}$ and can therefore be considered as
independent parameters.

For a 2D lattice of cavities, we assume the Bloch ansatz for the
polarization field, i.e.,
\begin{equation}
{\bf P}_{i}={\bf P}_{n \beta}=\exp (i {\bf k} \cdot {\bf R}_{n})
{\bf P}_{0 \beta} \label{eq:bloch}
\end{equation}
The cavity index $i$ becomes composite, $i \equiv n \beta$, where
$n$ enumerates the unit cell and $\beta$ the positions of
inequivalent cavities in the unit cell. Also, ${\bf R}_{n}$
denotes the lattice vectors and ${\bf k}=(k_{x},k_{y})$ is the
Bloch wavevector. By substituting Eq.~(\ref{eq:bloch}) into
Eq.~(\ref{eq:cde_eigen}) we finally obtain
\begin{equation}
\sum_{\beta'} \tilde{{\bf G}}_{\beta \beta'}(\omega_0,{\bf k})
{\bf P}_{0 \beta'}= \Omega {\bf P}_{0 \beta}
\label{eq:cde_eigen_periodic}
\end{equation}
where
\begin{equation}
\tilde{{\bf G}}_{\beta \beta'}(\omega_0, {\bf k}) = \sum_{n'} \exp
[i {\bf k} \cdot ({\bf R}_{n}-{\bf R}_{n'})] {\bf G}_{n \beta; n'
\beta'}(\omega_0). \label{eq:green_fourier}
\end{equation}
Solution of Eq.~(\ref{eq:cde_eigen_periodic}) provides the TE
frequency band structure of a periodic system of cavities.

\begin{figure}[h]
\centerline{\hbox{\includegraphics[width=3cm]{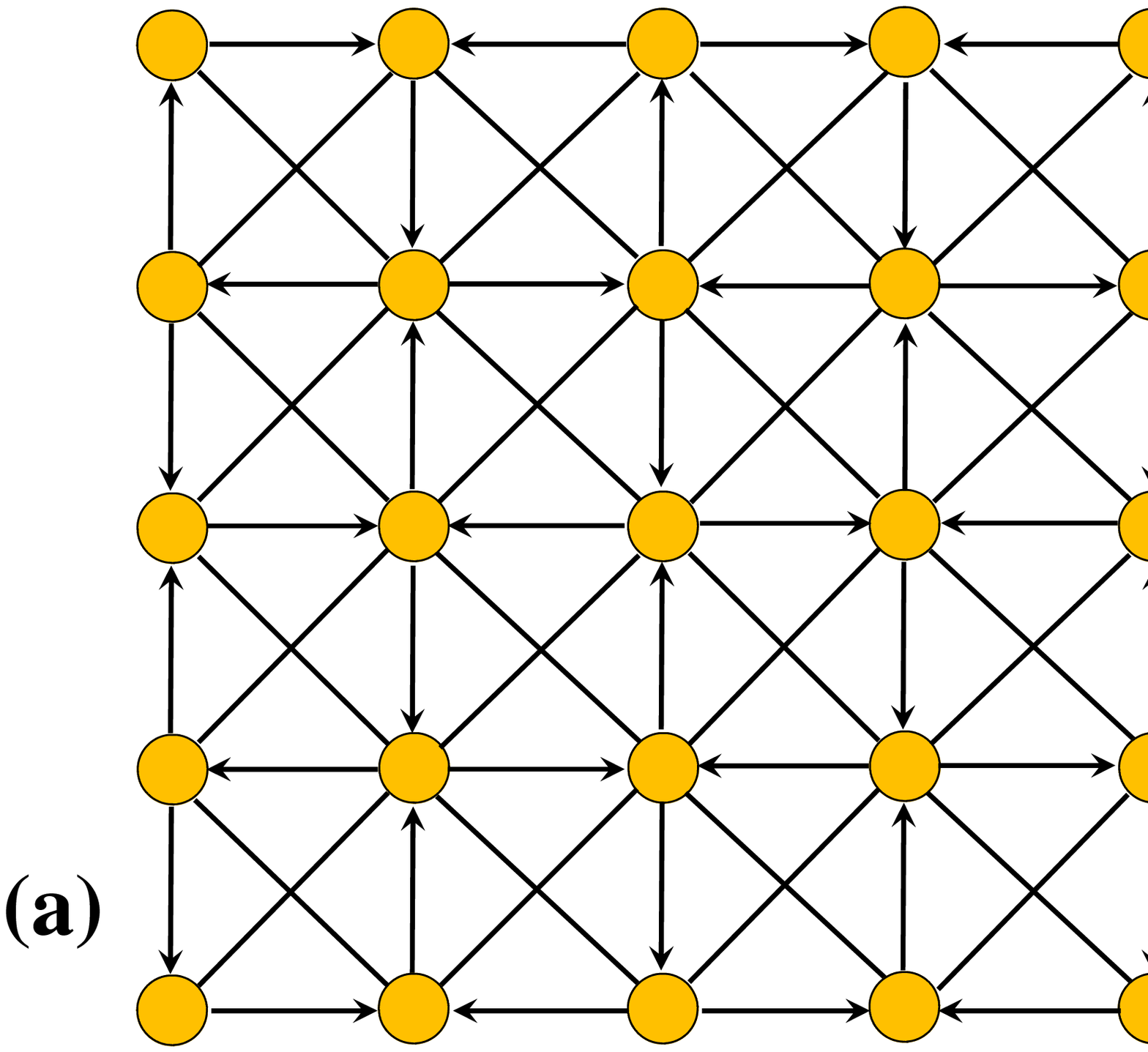}
\hbox{\includegraphics[width=6cm]{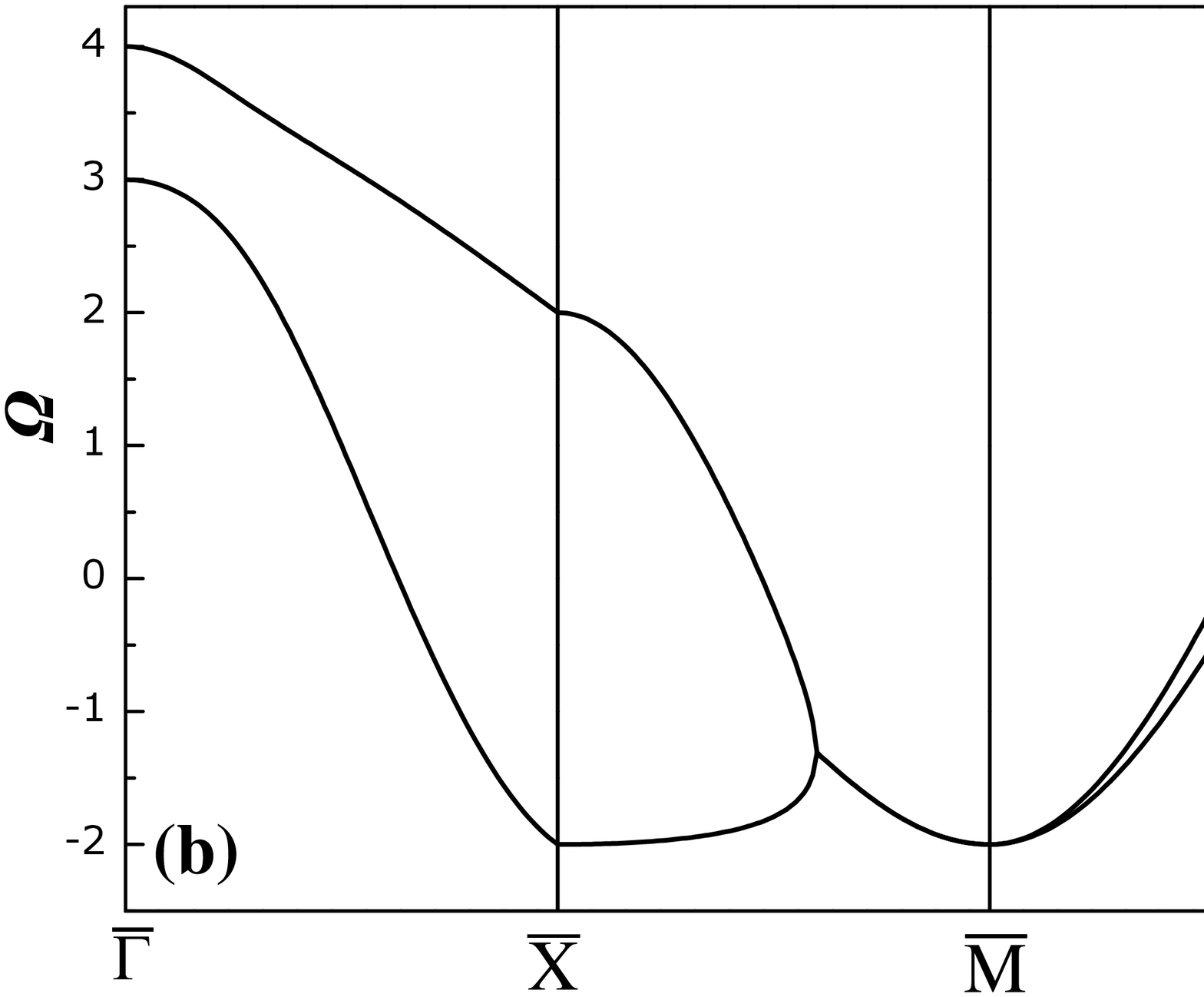} }}} \caption{(Color
online) (a) Square lattice of EM resonators connected with
metamaterial-based coupling elements. The arrows denote
nearest-neighbor hoppings whilst the solid lines
next-nearest-neighbor hoppings. The direction of the arrow shows
whether wave propagation in the coupling element is left- or
right-handed. (b) Frequency band structure corresponding to
lattice of the left panel for $t=t'=1$, $\phi=\pi/3$.}
\label{fig1}
\end{figure}

In order to seek for topological Bloch modes in a 2D lattice, we
need at least two distinct frequency bands. Since, TE modes
correspond to two degrees of freedom for the polarization field,
i.e., $(P_{x},P_{y})$ we may consider a 2D lattice with one cavity
per unit cell. Namely, we consider the square lattice of
Fig.~\ref{fig1}a where we consider nearest-neighbor (NN) and
next-nearest-neighbor (NNN) hoppings of the EM field among the
cavities. The NN hopping carries a nonzero phase $t \exp(\pm
\phi)$ whose signs are denoted by the arrows in Fig.~\ref{fig1}a.
The NNN hopping is denoted by $t'$. For the lattice of
Fig.~\ref{fig1}a, the Green's tensor $\tilde{{\bf G}}$ of
Eq.~(\ref{eq:green_fourier}) becomes
\begin{eqnarray}
\tilde{{\bf G}}&=& t [\cos\phi \cos(k_x \alpha/2) \cos(k_y
\alpha/2) - i \sin\phi \sin(k_x \alpha/2) \sin(k_y \alpha/2)]
\left(%
\begin{array}{cc}
3 & -1 \\
  -1 & 3 \\
\end{array}%
\right) \nonumber \\ &+&
2t'\left(%
\begin{array}{cc}
\cos (k_{y} \alpha) & 0 \\
  0 & \cos (k_{x} \alpha) \\
\end{array}%
\right) \label{eq:green_TE}
\end{eqnarray}

Fig.~\ref{fig1}b shows the frequency band structure derived from
Eq.~(\ref{eq:green_TE}) for $t=t'=1$, $\phi=\pi/3$. We observe
that at some point along the $\overline{X}\overline{M}$ symmetry
line the frequency bands coalesce into a single band. The point
beyond which the bands coalesce is an exceptional point and has
been observed in $\mathcal{PT}$-symmetric lattices. \cite{pt_sym}
In general, exceptional points emerging in parameter space are
associated with topological charge and geometric (Berry) phase.
\cite{ep_topol} The topological properties of an exceptional point
have been revealed by encircling it in parameter space \cite{hess}
as it was demonstrated in a microwave cavity experiment.
\cite{dembowksi} Although a proper theory of the topological
properties of the exceptional points in lattices is still lacking,
based on previous work \cite{ep_topol} we can indirectly assign
topological charge and geometric phase to the exceptional point
appearing in the frequency band structure of Fig.~\ref{fig1}b.

\normalsize

\subsection{TM modes}

\begin{figure}[h]
\vbox{\includegraphics[width=3cm]{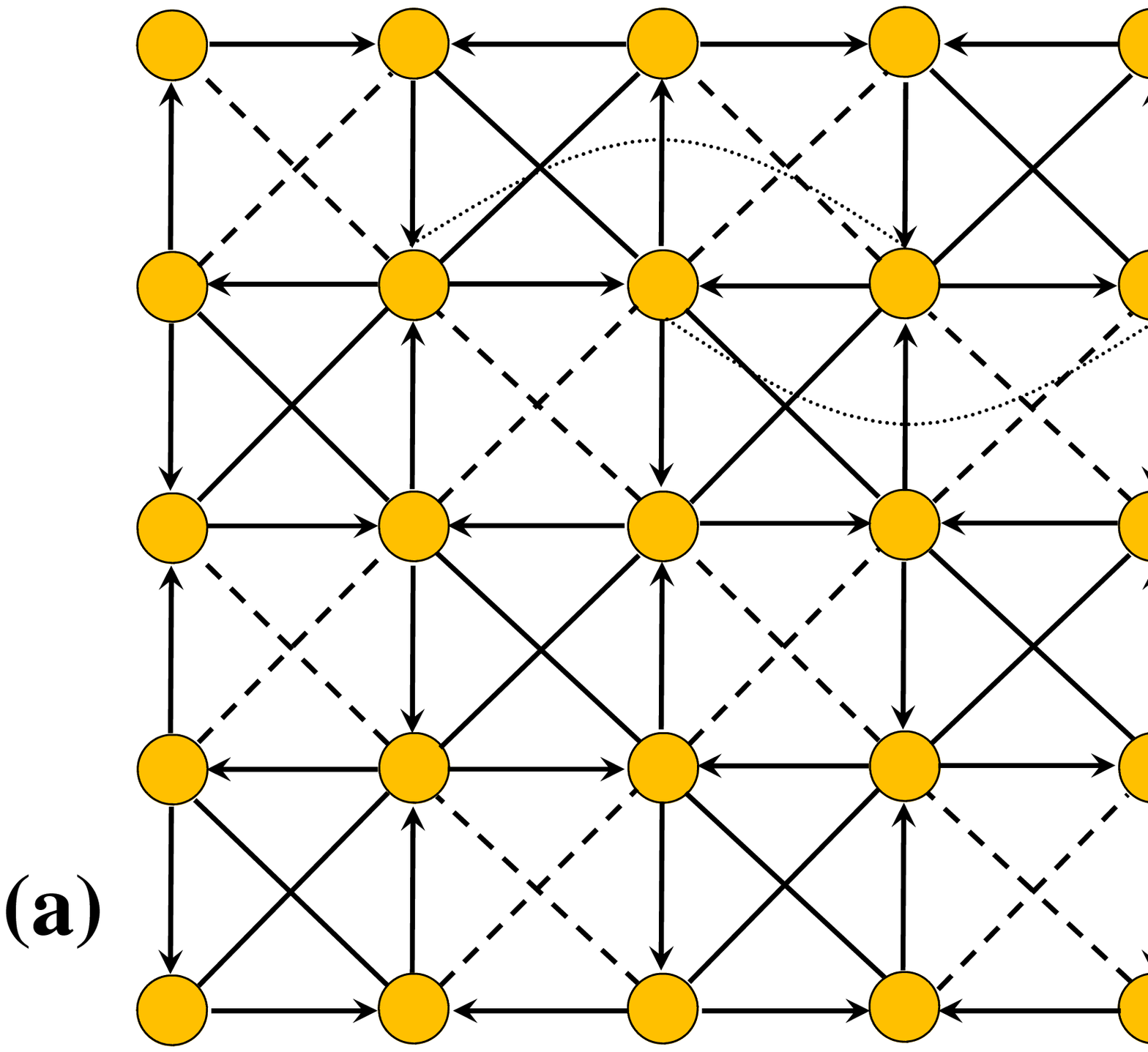}}
\vbox{\includegraphics[width=8cm]{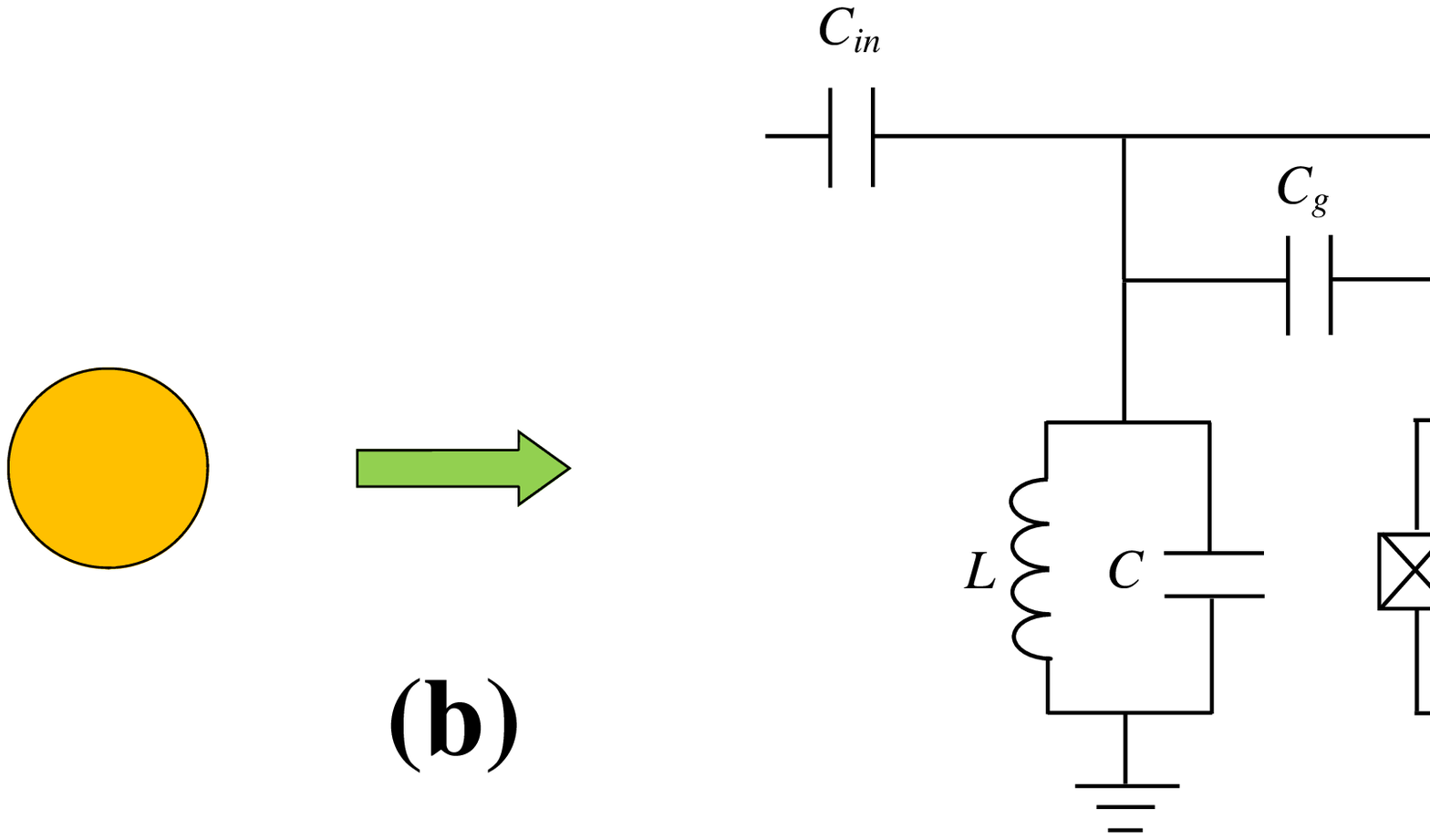}}
\vbox{\includegraphics[width=8cm]{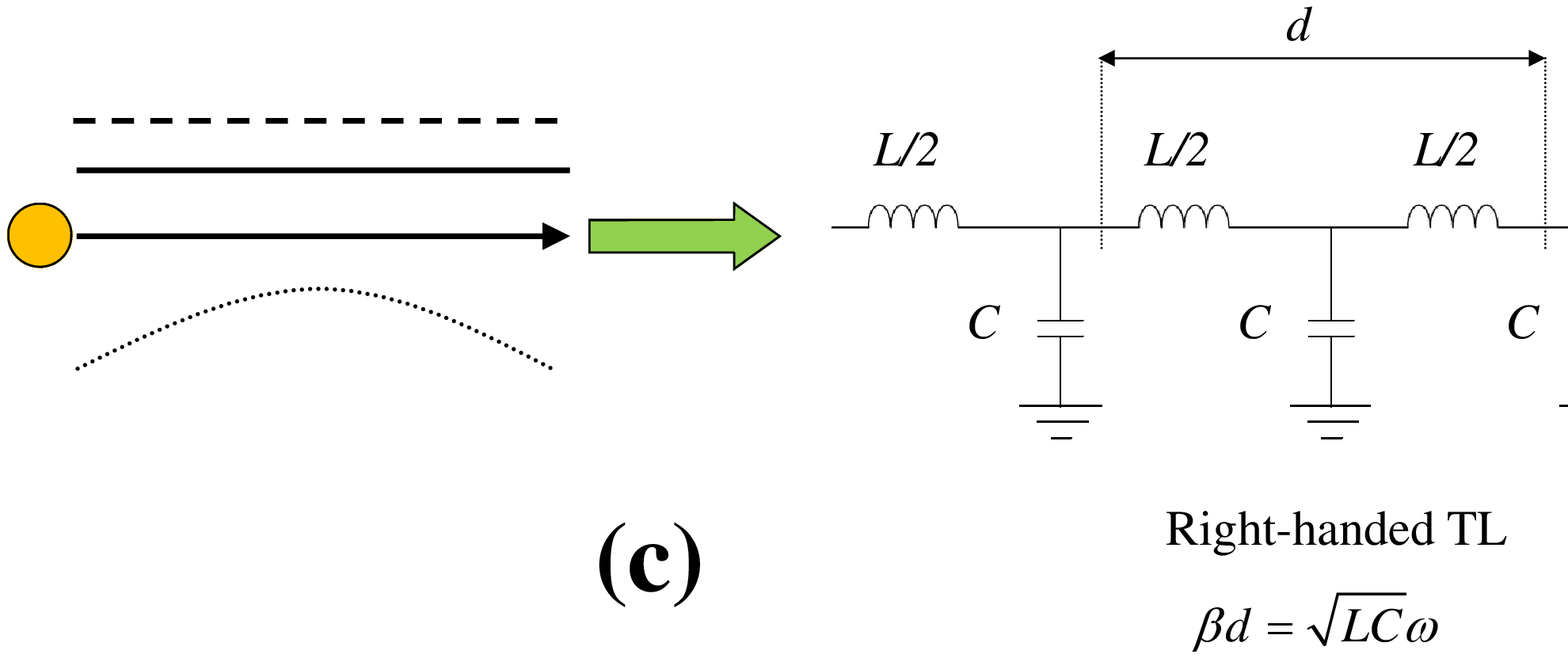}}
\vbox{\includegraphics[width=8cm]{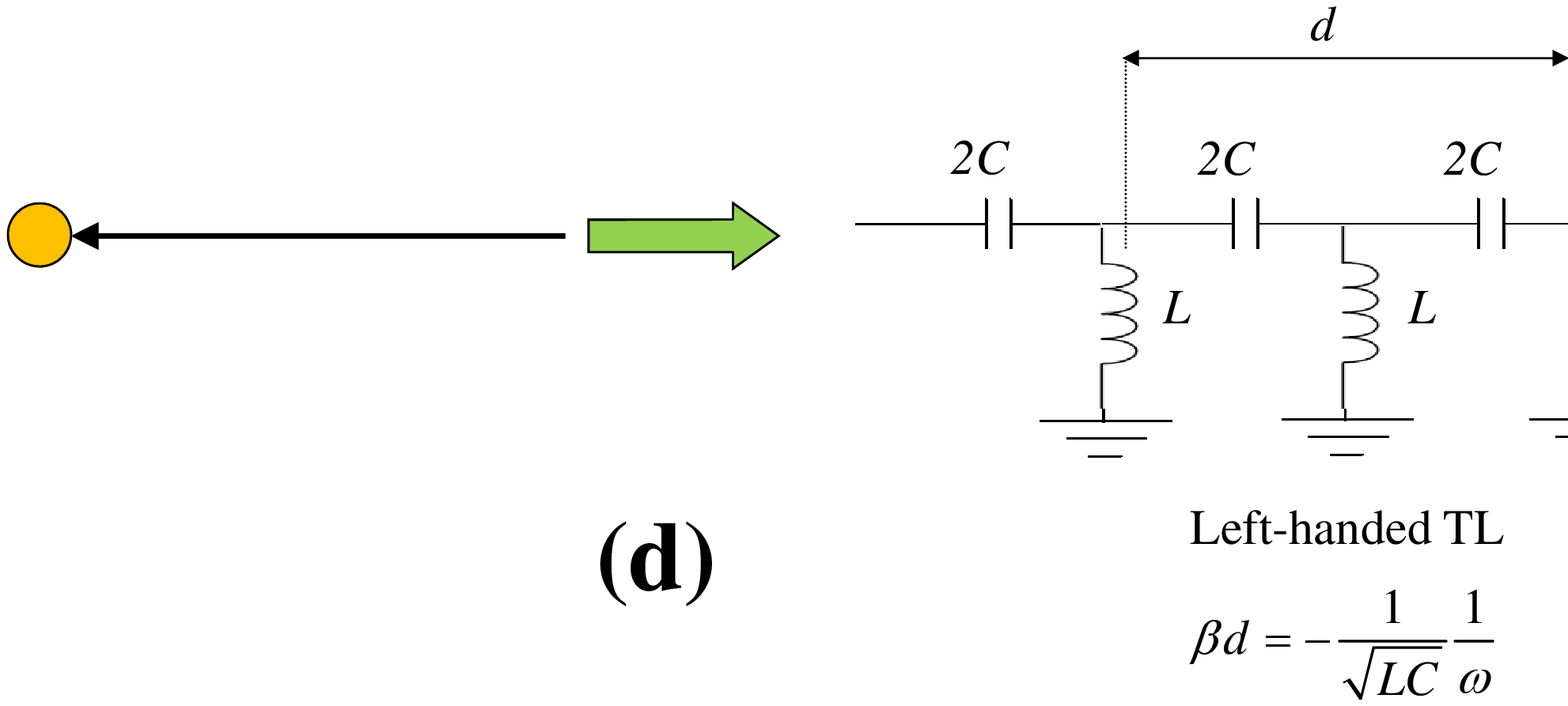} } \caption{(Color
online) (a) Checkerboard lattice of EM resonators connected with
metamaterial-based coupling elements. The arrows denote
nearest-neighbor hoppings whilst the solid and broken lines
next-nearest-neighbor hoppings. The direction of the arrow shows
whether wave propagation in the coupling element is left- or
right-handed. Two of the next-next-nearest neighbor hoppings are
shown as dotted arcs. (b) The EM resonator of the lattice is a
superconducting circuit QED system consisting of an {LC} resonator
coupled to Cooper pair box (CPB). Typical transmission line for
(c) right- ((d) left-) handed coupling elements along with the
corresponding dispersion relation.} \label{fig2}
\end{figure}

Next, we assume that the polarization at each dipole is oriented
in the $z$-axis. In this case, the eigenvalue problem of
Eq.~(\ref{eq:cde_eigen}) becomes scalar and
\begin{equation}
 G_{i i'}=
F q_{0}^{3} C(q_{\parallel} | r_{ii'}|) \simeq q_{0}^{3} F \exp(i
q_0 | r_{ii'}|)/ (q_0 |r_{ii'}|) = t_{ii'} \exp(i \phi_{ii'}).
\label{eq:green_tm}
\end{equation}
The same applies to
Eqs.~(\ref{eq:cde_eigen_periodic})-(\ref{eq:green_fourier}) and
the TM problem becomes equivalent to the electronic case. Since
the minimal model to have topological frequency bands is a
two-band model, we adopt the checkerboard lattice of
Ref.~\onlinecite{sun_prl} (see Fig.~\ref{fig2}a). Namely, apart
from considering NN and NNN hoppings as in Fig.~\ref{fig1}a, we
also consider next-next-nearest-neighbor (NNNN) hoppings (denoted
by the arcs in Fig.~\ref{fig2}a) with strength $t''$. The NN
hoppings are, again, complex, $t \exp(\pm \phi)$ where the sign is
denoted by the arrows in Fig.~\ref{fig2}a. The NNN hopping
strength is $t'_{1}$ ($t'_{2}$) if two sites are connected by a
solid (dashed) line. We note that in EM lattices such as those
considered here, a negative phase $-\phi$ can be easily achieved
when the cavities are connected e.g., by 1D left-handed
transmission lines  (LHTL), i.e. transmission lines supporting
backward-propagating waves where the phase velocity is opposite to
the group velocity. \cite{ele_1,caloz} Alternatively, the cavities
may be connected by waveguides loaded with a left-handed (LH)
metamaterial. Obviously, a positive phase $+ \phi$ can be achieved
by similar means [right-handed transmission lines (RHTLs)].

\small
\normalsize

\begin{figure}[h]
\vbox{\includegraphics[width=6cm]{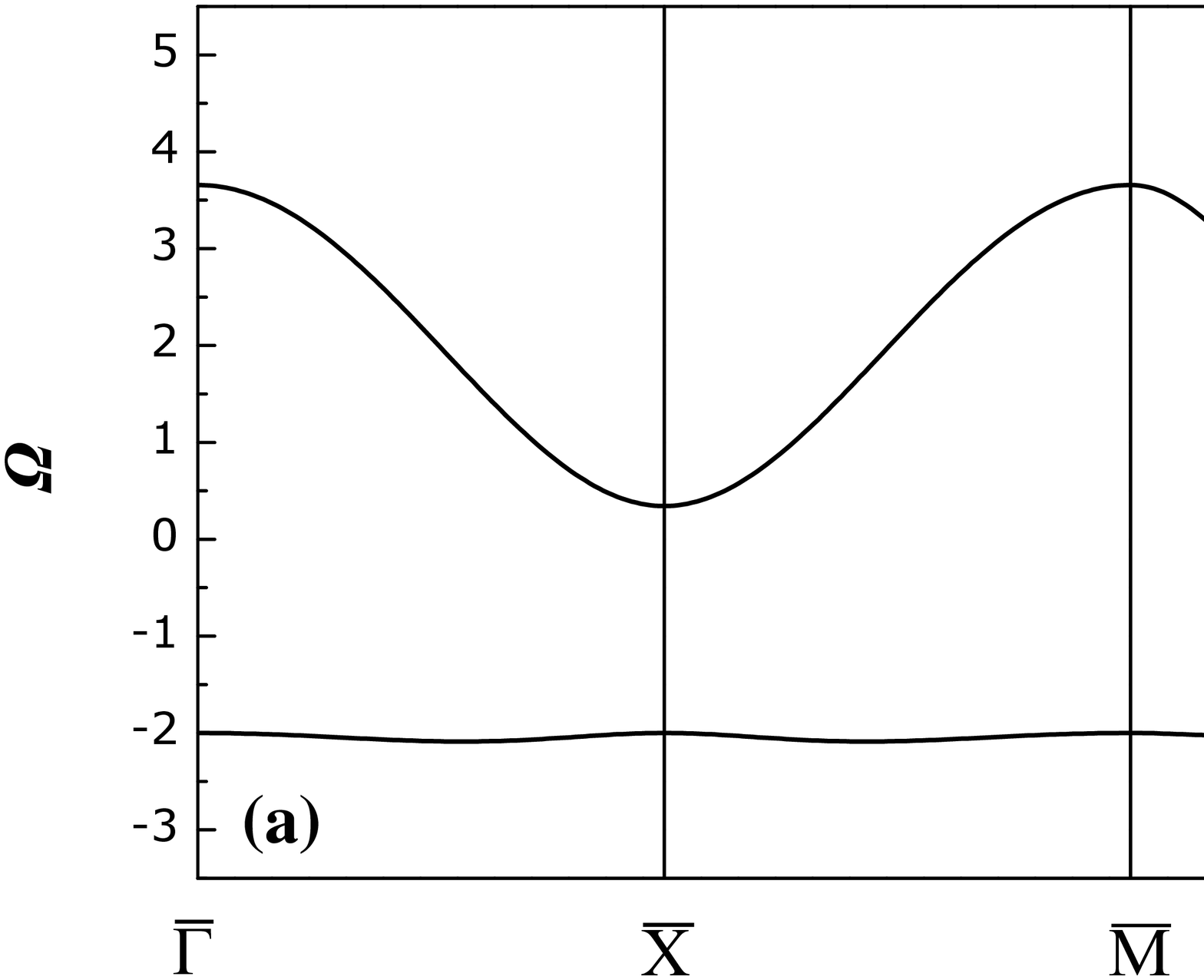}}
\vbox{\includegraphics[width=6cm]{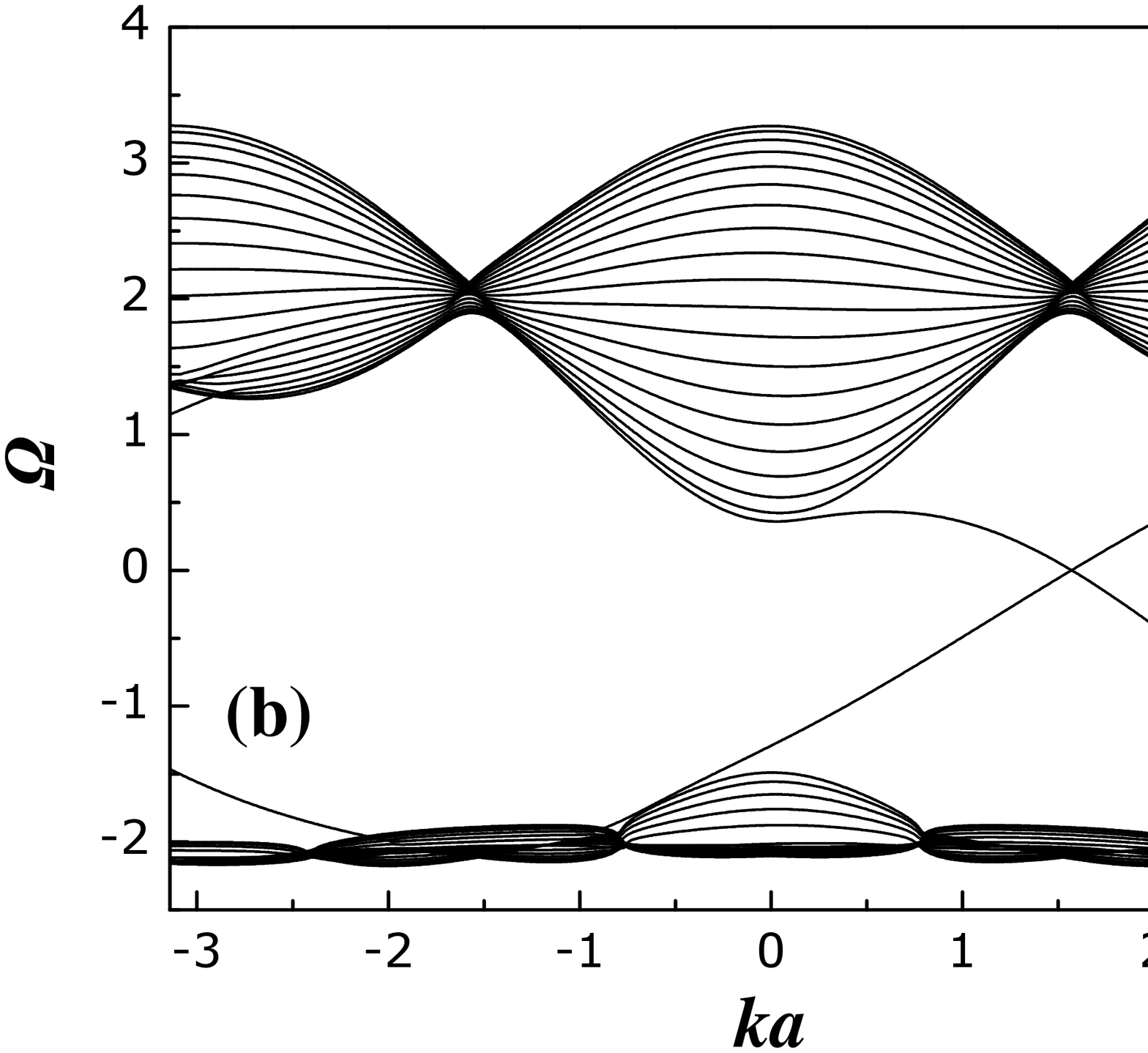}} \caption{(a)
Frequency band structure for the infinite checkerboard lattice of
Fig.~\ref{fig2} and (b) the frequency band structure for a finite
slab consisting 20 unit planes [parameters: $t=1$, $\phi=\pi/4$,
$t'_{1}=-t'_{2}=1/(2+\sqrt{2})$, $t''=1/(2+2\sqrt{2})$] }
\label{fig3}
\end{figure}

For the checkerboard lattice of Fig.~\ref{fig2}a, the Green's
tensor ${\tilde {\bf G}}_{\beta \beta'}$ becomes
\begin{equation}
\tilde{{\bf G}}= \left(%
\begin{array}{cc}
G_{11}
 & G_{12} \\
 G_{21}  &
 G_{22} \\
\end{array}%
\right) \label{eq:green_TM_fourier}
\end{equation}
where
\begin{eqnarray}
G_{11}=2 t'_{1} \cos(k_{x} \alpha) +  2 t'_{2} \cos(k_{y} \alpha)
 + 4t'' \cos(k_{x} \alpha) \cos(k_{y} \alpha) \nonumber \\
G_{12}=G^{*}_{21}=4t \cos \phi \cos(k_{x} \alpha/2) \cos(k_{y}
\alpha/2) -4 i t \sin \phi \sin(k_{x} \alpha/2) \nonumber \\
G_{22}=2 t'_{2} \cos(k_{x} \alpha) +  2 t'_{1} \cos(k_{y} \alpha)
 + 4t'' \cos(k_{x} \alpha) \cos(k_{y} \alpha)
 \label{eq:green_elements}
 \end{eqnarray}
Atomic lattices with exotic hoppings such as those considered here
have been used for simulating fractional Quantum Hall effect
(FQHE) states at zero magnetic field as nearly flat topological
bands can emerge which simulate the Landau levels associated with
a uniform magnetic field. \cite{neupert,wang,sheng} By taking as
TB parameters, \cite{sun_prl} $t=1$, $\phi=\pi/4$,
$t'_{1}=-t'_{2}=1/(2+\sqrt{2})$, $t''=1/(2+2\sqrt{2})$], a nearly
flat band emerges as it is evident from the frequency band
structure of Fig.~\ref{fig3}a. Based on the equivalence of the
Green's tensor of Eqs.~(\ref{eq:green_TM_fourier}) and
(\ref{eq:green_elements}) with the Hamiltonian of the electronic
problem, \cite{sun_prl} each of the two bands of Fig.~\ref{fig3}a
carries a Chern number $\pm 1$. The topological nature of the
frequency bands of the lattice of Fig.~\ref{fig2}a is also
manifested by the emergence of one-way bands (the photonic
counterpart of the electron chiral edge states
\cite{haldane,one_way}) in the frequency band structure for a slab
geometry of Fig.~\ref{fig3}b.

\small
\normalsize

The occurrence of topological properties such as the exceptional
point in Fig.~\ref{fig1}b and the one-way modes in Fig.~\ref{fig3}
are a result the synthetic gauge field which is generated by the
geometry of the metamaterial-based coupling elements (formation of
closed flux loops of the phase of the EM field).

\section{Simulation of the FQHE}

Having established a nearly flat topological frequency band of the
EM field for the lattice of Fig.~\ref{fig2}a, we are able to
design a system for creating an EM analog of the FQHE. The most
natural choice would be to consider a coupled cavity array (CCW)
wherein polaritons propagate through a hopping mechanism (as in
our case) and interact strongly with the reservoir of modes when
they reside within the cavity. \cite{ccw,zhou} FQHE with magnetic
field can also be simulated by atoms confined in a 2D CCW.
\cite{angelakis} As stated above, the EM lattice of
Fig.~\ref{fig2}a can be realized in the laboratory as a network of
transmission lines (TLs) where hoppings with positive (negative)
phase can be realized with RHTLs (LHTLs) as shown in
Fig.~\ref{fig2}c (Fig.~\ref{fig2}d). This means that the
topological bands lie in the GHz regime. Therefore, in order to
simulate the FQHE for microwave photons we need to implement a
cavity QED scheme in this regime. This can be achieved by
considering a superconducting-circuit cavity QED system
\cite{wallraf,blais} consisting of a Cooper-pair box (CPB) coupled
to a TL resonator (see the equivalent circuit of
Fig.~\ref{fig2}b). The CPB operates as an artificial atom
(two-level system) \cite{bouchiat,clarke} and couples to the
microwave photons of a superconducting TL resonator which plays
the role of an on-chip cavity reservoir. The microwave response of
superconducting circuit cavity QED system is described by the
Jaynes-Cummings Hamiltonian \cite{wallraf,blais}
\begin{eqnarray}
H^{JC}&=&\hbar \omega_{r} (a_{i}^{+} a_{i}
+1/2)-\frac{1}{2}(E_{el} \sigma_{x}^{i} + E_{J} \sigma_{z}^{i})
\\ \nonumber &+& \hbar g (a_{i}^{+} \sigma_{i}^{-} + a_{i} \sigma_{i}^{+})
\label{eq:h_jc}
\end{eqnarray}
where $\omega_{r}=1/\sqrt{LC}$ is the frequency of the
superconducting resonator, $a_{i}^{+}$ ($a_{i}$) creates
(annihilates) a microwave photon in the TL resonator (cavity),
$\sigma_{i}^{+}$ ($\sigma_{i}$) creates (annihilates) an
excitation in the CPB, $g$ is the coupling parameter between the
CPB and the TL resonator, $E_{el}$ is the electrostatic energy and
$E_{J}=E_{J,max}\cos(\pi \Phi_{b})$ the Josephson energy of the
CPB. $\Phi_{b}=\Phi/ \Phi_{0}$ is a flux bias applied by a coil to
the CPB and controls the Josephson energy $E_{J}$.

A superconducting circuit cavity QED system where microwave
photons propagate in the lattice of Fig.~\ref{fig2}a is described
by a Jaynes-Cummings-Hubbard Hamiltonian of the form \cite{makin}
\begin{equation}
H^{JCH}=\sum_{i} H^{JC}_{i} + H^{TB} \label{eq:h_jch}
\end{equation}
where $H^{TB}$ is the tight-binding form of the Hamiltonian of the
microwave photons propagating within the checkerboard lattice of
Fig.~\ref{fig2}a, i.e.,
\begin{eqnarray}
H^{TB}=&-&t \sum_{\langle i,j \rangle} \exp(i \phi_{ij})
(a_{i}^{+} a_{j} + H.c.) -\sum_{\langle\langle i,j \rangle\rangle}
t'_{ij} (a_{i}^{+}a_{j} + H.c.) \nonumber \\ &-& t''
\sum_{\langle\langle\langle i,j \rangle\rangle\rangle} (a_{i}^{+}
a_{j} + H.c.) \label{eq:h_tb}
\end{eqnarray}
which is the direct-space representation of the Green's tensor of
Eq.~(\ref{eq:green_TM_fourier}). An important ingredient which
gives rise to the FQHE is the presence of repulsive interactions
among the microwave photons and is inherently present in
Eq.~(\ref{eq:h_jch}) as photon blockade. \cite{makin} The latter
phenomenon has been recently observed experimentally in the GHz
regime for superconducting circuit cavity QED systems such as the
one considered here (CPB + TL resonator). \cite{lang,hoffman} The
different FQHE phases can be calculated by direct-diagonalization
of the Hamiltonian of Eq.~(\ref{eq:h_jch}). We note that the
proposed quantum simulator for the FQHE differs fundamentally with
previous proposals \cite{koch,hafezi} since it essentially
constitutes a passive design requiring no externally applied
electric or magnetic fields.

Some typical values of the Hamiltonian of Eq.~(\ref{eq:h_jch})
are: \cite{wallraf} $\omega_{r}= 38$~GHz, $E_{J,max}=8 GHz$,
$E_{C}=5.2GHz$, $g \approx 0.314$~GHz. The latter parameter, $g$,
is much larger than the loss rate of the TL resonator ($\sim
0.005$~GHz) and the decoherence rate of the CPB ($\sim
0.004$~GHz). The frequency $\omega_{r}$ of the TL resonator should
fall within the operating bandwidth of the LH- and RHTLs. In
Fig.~\ref{fig2}c and Fig.~\ref{fig2}d we have considered ideal TL
which have infinite bandwidth. However, actual LH- and RHTLs have
very large bandwidth which is a distinctive feature of nonresonant
metamaterials compared to the resonant ones. \cite{caloz} Lastly,
if the superconducting QED chip has a thickness of 1mm, the
coupling TLs (RH or LH) have 10mm length and the TL resonator
covers an area of $30{\rm mm}^{2}$, for the given resonator
frequency (38GHz), the hopping strength $t$ is about 0.5GHz which
is also significantly larger than both the TL loss and CPB
decoherence rates.

In order to probe experimentally the FQHE with the proposed
structure one needs to create the phase diagram of the spectrum
gap between the FQHE ground-state manifold and the lowest excited
states, as a function of the coupling parameters $g$ for NN and
NNN hopping when the latter lie in the photon blockade regime.
Generally speaking, in the FQHE state the spectral gap assumes
much larger values than in superfluid and solid phases.
\cite{wang} The frequencies of the ground and excited states (and
thus their corresponding gaps) can be measured by microwave
transmission experiments.

\section{Conclusion}
In conclusion, we have shown that topological frequency bands
emerge in 2D electromagnetic lattices of metamaterial components
in the absence of an applied magnetic field. The topological
nature of the corresponding band structures gives rise to
significant phenomena such as one-way waveguiding and coalescence
of EM modes. The above lattices can be the basis for realizing a
simulator for the FQHE based on superconducting transmission lines
and circuit cavity QED systems.


\begin{thebibliography}{}
\bibitem{haldane} F.~D.~M.~Haldane and S.~Raghu, \prl {\bf 100},
013904 (2008); {\it ibid}, \pra {\bf 78}, 033834 (2008).
\bibitem{one_way} Z.~Wang, Y.~D.~Chong, J.~D.~Joannopoulos, and M.~Solja\v{c}i\'{c}, \prl
{\bf 100}, 013905 (2008); {\it ibid}, Nature (London) {\bf 461},
772 (2009); Z.~Yu, G.~Veronis, Z.~Wang, and S.~Fan, \prl {\bf
100}, 023902 (2008); H.~Takeda and S.~John, \pra {\bf 78}, 023804
(2008); D.~Han, Y.~Lai, J.~Zi, Z.~Q.~Zhang, and C.~T.~Chan, \prl
{\bf 102}, 123904 (2009); X.~Ao, Z.~Lin, and C.~T.~Chan, \prb {\bf
80}, 033105 (2009); M.~Onoda and T.~Ochiai, \prl {\bf 103}, 033903
(2009); T.~Ochiai and M.~Onoda, \prb {\bf 80}, 155103 (2009);
R.~Shen, L.~B.~Shao, B.~Wang, and D.~Y.~Xing, \prb {\bf 81},
041410(R) (2010);
 Y.~Poo, R.~X.~Wu, Z.~Lin, Y.~Yang, and C.~T.~Chan,
\prl {\bf 106}, 093903 (2011).
\bibitem{yannop_rashba} V.~Yannopapas, \prb {\bf 83}, 113101
(2011).
\bibitem{ti_papers} L.~Fu, C.~L.~Kane, and E.~J.~Mele, \prl {\bf
98}, 106803 (2007); L.~Fu and C.~L.~Kane, \prb {\bf 76}, 045302
(2007); M.~Z.~Hasan and C.~L.~Kane, \rmp {\bf 82}, 3045 (2010).
\bibitem{mele_2005} C.~L.~Kane and E.~J.~Mele, \prl {\bf 95},
226801 (2005); {\it ibid}, \prl {\bf 95}, 146802 (2005).
\bibitem{chen_pti} W.-~J.~Chen, Z.~H.~Hang, J.-~W.~Dong, X.~Xiao,
H.-~Z.~Wang, and C.~T.~Chan, \prl {\bf 107}, 023901 (2011).
\bibitem{zhang_pti} W.~Zhong and X.~Zhang, Opt.~Express {\bf 19},
13738 (2011).
\bibitem{yannop_pti} V.~Yannopapas, \prb {\bf 84}, 195126 (2011).
\bibitem{cde} E.~M.~Purcell and C.~R.~Pennypacker, Astrophys.~J.
{\bf 186}, 705 (1973).
\bibitem{pt_sym} K.~G.~Makris, R.~El-Ganainy, D.~N.~Christodoulides,
and Z.~H.~Musslimani, \prl {\bf 100}, 103904 (2008); S.~Klaiman,
U.~G\"{u}nther, and N.~Moiseyev, \prl {\bf 101}, 080402 (2008);
S.~Longhi, \prl {\bf 103}, 123601 (2009); M.~Botey, R.~Herrero,
and K.~Staliunas, \pra {\bf 82}, 013828 (2010).
\bibitem{ep_topol} U.~G\"{u}nther, I.~Rotter, and B.~F.~Samsonov,
J.~Phys.~A:~Math.~Theor. {\bf 40}, 8815 (2007); F.~Keck,
H.~J.~Korsch, and S.~Mossman, J.~Phys.~A:~Math.~Gen. {\bf 36},
2125 (2003); A.~I.~Nesterov and F.~Aceves~de~la~Cruz,
J.~Phys.~A:~Math.~Theor. {\bf 41}, 485304 (2008).
\bibitem{hess} W.~D.~Hess, Eur.~Phys.~J.~D {\bf 7}, 1 (1999).
\bibitem{dembowksi} C.~Dembowski, H.-~D.~Gr\"{a}f, H.~L.~Harney,
A.~Heine, W.~D.~Heiss, H.~Rehfeld, and A.~Richter, \prl {\bf 86},
787 (2001).
\bibitem{sun_prl} K.~Sun, Z.~Gu, H.~Katsura, and S.~Das~Sarma, \prl
{\bf 106}, 236803 (2011).
\bibitem{ele_1} A.~K.~Iyer and G.~V.~Eleftheriades, {\it Negative
Refraction Metamaterials: Fundamental Principles and
Applications}, edited by G.~V.~Eleftheriades and K.~G.~Balmain
(Wiley-IEEE Press, New York, 2005) p.~1
\bibitem{caloz} C.~Caloz and T.~Itoh, {\it Electromagnetic
Metamaterials: Transmission Line Theory and Microwave
Applications} (Wiley-IEEE Press, New Jersey, 2006).
\bibitem{neupert} T.~Neupert, L.~Santos, C.~Chamon, and C.~Mudry,
\prl {\bf 106}, 236804 (2011).
\bibitem{wang} Y.-~F.~Wang, Z.-~C.~Gu, C.-~D.~Gong, and
D.~N.~Sheng, \prl {\bf 107}, 146803 (2011).
\bibitem{sheng} D.~N.~Sheng, Z.-~C.~Gu, K.~Sun, and L.~Sheng,
Nature~Commun., DOI:10.1038/ncomms1380, (2011).
\bibitem{ccw} A.~D.~Greentree, C.~Tahan, J.~H.~Cole, and
L.~C.~L.~Hollenberg, Nature~Phys. {\bf 2}, 856 (2006);
M.~J.~Hartmann, F.~G.~S.~L.~Brand\~{a}o, and M.~B.~Plenio,
Nature~Phys. {\bf 2}, 849 (2006); D.~G.~Angelakis, M.~F.~Santos,
and S.~Bose, \pra {\bf 76}, 031805 (2007); M.~J.~Hartmann,
F.~G.~S.~L.~Brand\~{a}o, and M.~B.~Plenio, Laser~Photon.~Rev. {\bf
2}, 527 (2008).
\bibitem{zhou} L.~Zhou, Y.~B.~Gao, Z.~Song, and C.~P.~Sun, \pra {\bf 77},
013831 (2008).
\bibitem{angelakis} J.~Cho, D.~G.~Angelakis, and S.~Bose, \prl
{\bf 101}, 246809 (2008).
\bibitem{wallraf} A.~Wallraff, D.~I.~Schuster, A.~Blais, ~L.~Frunzio,
R.-~S.~Huang, J.~Majer, S.~Kumar, S.~M.~Girvin, and
R.~J.~Schoelkopf, Nature (London) {\bf 431}, 162 (2004).
\bibitem{blais} A.~Blais, R.-~S.~Huang, A.~Wallraff, S.~M.~Girvin,
and R.~J.~Schoelkopf, \pra {\bf 69}, 062320 (2004).
\bibitem{bouchiat} V.~Bouchiat, D.~Vion, P.~Joyez, D.~Esteve, and
M.~H.~Devoret, Phys.~Scripta {\bf 76}, 165 (1998).
\bibitem{clarke} J.~Clarke and F.~K.~Wilhelm, Nature (London) {\bf
453} 1031 (2008).
\bibitem{makin} M.~I.~Makin, J.~H.~Cole, C.~Tahan,
L.~C.~L.~Hollenberg, and A.~D.~Greentree, \pra {\bf 77}, 053819
(2008).
\bibitem{lang} C.~Lang, D.~Bozyigit, C.~Eichler, L.~Steffen,
J.~M.~Fink, A.~A.~Abdumalikov~Jr., M.~Baur, S.~Filipp,
M.~P.~da~Silva, A.~Blais, and A.~Wallraff, \prl {\bf 106}, 243601
(2011).
\bibitem{hoffman} A.~J.~Hoffman, S.~J.~Srinivasan, S.~Schmidt,
L.~Spietz, J.~Aumentado, H.~E.~T\"{u}reci, and A.~A.~Houck, \prl
{\bf 107}, 053602 (2011).
\bibitem{koch} J.~Koch, A.~A.~Houck, K.~L.~Hur, and S.~M.~Girvin,
\pra {\bf 82}, 043811 (2010); A.~Nunnenkamp, J.~Koch, and
S.~M.~Girvin, New~J.~Phys. {\bf 13}, 095008 (2011).
\bibitem{hafezi} M.~Hafezi, E.~A.~Demler, M.~D.~Lukin, and
J.~M.~Taylor, Nature~Phys. {\bf 7}, 907 (2011).
\end{thebibliography}
\end{document}